\documentclass[10pt,showpacs,preprintnumbers,amsmath,amssymb]{revtex4}

\usepackage{epsf}
\usepackage{graphicx}  
\usepackage{dcolumn}   
\usepackage{bm}        
\newcommand{\be}{\begin{equation}}
\newcommand{\en}{\end{equation}}
\newcommand{\ba}{\begin{eqnarray}}
\newcommand{\ea}{\end{eqnarray}}

\begin{document}

\title{Does the Cosmos have two times? Multi-time and cosmic acceleration}

 \author{ Hongsheng Zhang\footnote{Electronic address: hongsheng@shnu.edu.cn} }
  \affiliation{ Shanghai United Center for Astrophysics (SUCA),
 Shanghai Normal University, 100 Guilin Road, Shanghai 200234, P.R.China }
 \date{ \today}

\begin{abstract}
We put forward a multi-time theory,
in  frame of which the
cosmic acceleration is a natural phenomenon without
cosmological constant or anything like that. The main point of this
theory is that each of the gravity interaction and electromagnetic
interaction has its own time, respectively. Also we give a concrete
model of this theory which can exactly simulate $\Lambda$CDM. Further
 we discuss the possible observations which may improve this theory in
the future.

\end{abstract}


\maketitle

\begin{figure}
\centering
\includegraphics[height=7cm,  angle=0]{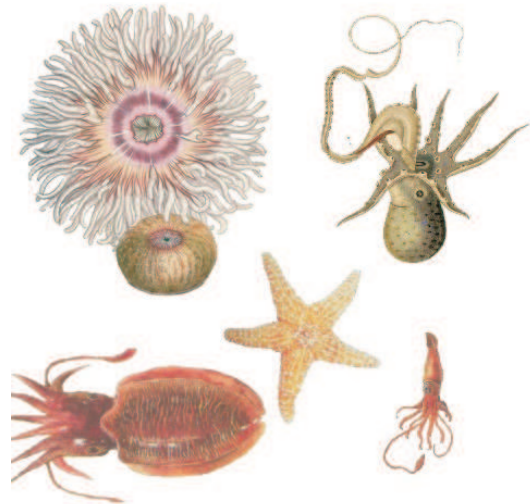}
\caption{The concept of space roots in the existence of rigid bodies. In a world of mollusca, how can they get the concept of
distance?}
\label{mollusca}
\end{figure}

  Time is one of the most ordinary and profound concept, for the
  public and for scientists. From the ancient times, numerous
  philosophers, humanists and poets offered creative ideas about
  time one after another. However, the following point is persisted
  in the modern science: A quantity which can not be measured
  quantitatively is not qualified as the object of science. According to this rule, ``space", as
  an object of science, founds on the rigid body. We can imagine a world of mollusca.
  Creatures living in such a world can never quantitatively measure a
  distance, as shown in Fig \ref{mollusca}. In fact, it is not necessary for them to develop the
  notions such as metric, or metric
  space, though they can develop topology.

     Similarly, the scientific concept of time founds on periodic
     motion. Time, in everyday life, may just denote event sequence,
     or psychological time, such as one car goes after another. When we see the cars go one after
     another, the ink disperses in water, or the galaxies are recessing, we know that time is flowing. If we
     just stop on ``event sequence", we can never  quantitatively
     measure an interval of time (Maybe we can develop a subject like time
     topology). We have to refer to a motion with repetition property.
     Fortunately, the Nature presents a type of motion which can help us
     to quantitatively measure the interval of time: It is  periodic motion.
     To measure the size of a space, we need to put a ruler (rigid body) between two extremes one
     time after another to fill the space, with an accuracy to the length of the
     ruler itself. To measure an interval of time, we need to observe how
     many cycles undergone in such an interval, with an accuracy to the period of the
     cycle itself. Hence, the periodic motion is necessary to
     quantitatively measure an interval of time.

    In the case of space measurement, we can freely shift the same ruler back
    and forth, which ensures that the space interval is gauged by the ruler
    into segments with the same length. However, the case of measurement
    of time is very different because the crucial property of time:
    Irreversibility, which makes that the idea
   of $equal$ of two periods one after the other is not as clear as the first
    sight.  We can measure an interval of time by a
    periodic motion. But we can not verify that the period of this
    cycle equals the last one. Philosophers and mathematicians noticed this point many
    years ago. Locke was aware of that ``no two parts of duration can be certainly known to be equal." \cite{locke}
     Leibniz wrote:`` our measure of time would be more accurate if
    we could keep one day past to compare with days to come, as we
    keep the measures of space." (translated by the
    author)\cite{leib}. Poincare also mentioned this point \cite{poincare}.
    He said, ``...2. There is no absolute time.
   When we say that two periods are equal, the statement has no meaning,
  and can only acquire a meaning by a convention.
   3. Not only have we no direct intuition of the equality of two periods,
   but we have not even direct intuition of the simultaneity of
   two events occurring in two different places..." The Poincare's point 2 is essentially the same as of
   Locke and Leibnitz's. Both of them stress that it is impossible to compare the successive time
   segments. We know that the Poincare's point 3
  triggered special relativity. On the contrary, the point 2 never get
  enough concerns. This problem highlights the irreversibility of time.
  For the case of space, we can take the $same$ ruler to measure the
  length of the different segments of a line by shifting the ruler
  freely, back and forth. But a time interval passed, and passed for
  ever. No clock can confirm that the last second is equal to this
  second. We say they are equal is only, in fact, a convention, or a stipulation.

   If we can not know whether the two successive time segments are
   equal indeed, or it is just an improper question, a concession is to
   find a  natural convention for this equality. An isolated periodic system
   without dissipation (There is no energy transfer between macroscopical
   motion and microscopic motion.) seems a natural candidate, that is, we stipulate
  that the time passed in every period is equal if the system of periodic
  motion is in isolated state without dissipation. We call a system satisfying these conditions a
  perfect system to measure time (PSMT). Approximately, the earth-sun
  system, earth-moon system, a mechanic clock with spring,
  are approximately PSMT, and have been used to
  measure time for many years. Sure, any realistic  system is not rigorously isolated
  and has more or less dispassion. But this does not prevent us from
  acquiring the concept of quantitative time, as we obtain the concept of metric
  space without perfect rigid body.

   If we have a PSMT, an immediately subsequent question whether  all PSMTs be
   synchronised.   To answer this question we need to study the mechanism of the PSMTs.
    There are several different types of periodic motions which are commanded by different interactions.
     In spite of  developments in unite theory in physics, the 4 (or 3) interactions we
 know today are inherently different all the same, especially the
 gravity is remarkably different from other three gauge
 interactions. In as much the interactions are intrinsically different
 we have no transcendental reason to assure PSMTs controlled by different
 interactions can be synchronised for ever, which means the time
 velocities may different in systems controlled by different
 interactions.

\begin{figure}
\centering
\includegraphics[height=8cm, angle=0]{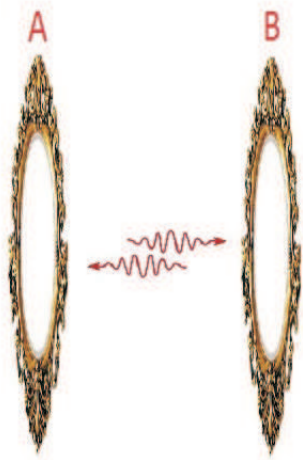}
\caption{A simplest light clock. A photon travels back and forth between two parallel mirrors.}
\label{light clock}
\end{figure}

  In a periodic system controlled by one
  interaction, we have a natural convention, that is, time flows equably in a PSMT.
    But, how about the conventions of the systems controlled by different interactions?
  If we apply our natural conventions to two different system, one
  is  controlled by gravity and the other is controlled by
  electromagnetic interaction, we are not sure whether the two conventions are
  identified.   This idea seems delicate so here we explain it
 explicitly by an example. We have two systems : the earth-sun
 system and a simplest light clock. The light clock is composed by
 two parallel mirrors, A and B, whose reflectivity is 100\%  and a photon
 travels back and forth between the two mirrors, as shown in fig \ref{light clock}. The light ray
 is  vertical to the mirrors. The time that this photon starts from mirror A
 to mirror B and then returns back to A we call a ``tick". We arrange the two
 mirrors in appropriate distance to ensure the light clock passes $10^{16}$ ticks when the earth rotates a circle
 around the sun. But in the
 interval that the earth rotating the sun next circle the light clock passes
 $10^{16}+1$ ticks. We suppose both of the two systems are at
 perfect isolated states. So under this condition we have to accept that
 the notion of time derived from periodic system of gravity is
 $inherently$ different from notion of time derived from periodic system of
 electromagnetism,  because we have no transcendental reason
 to infer that the convention in the first system is  superior than that of the second or vice
 versa. We have to admit that there are two inherently different
 times.

  Though philosophers and mathematicians such as Locke, Leibniz and Poincare have sensed
  the problem of equability of successive time segments, we are indifferent to the time conventions of
  different systems in concision sciences up to now.  We don't
   know how to set up a concrete scientific theory to describe it  because of the deficiency of experimental data. The
  recent cosmological observation, which implies the universe expands more and
   more faster,  offers an operative effect to test whether the conventions in different
   PSMTs are the same or not.

   First we display a toy model to show how the multi-time theory
   works in cosmology. In this toy model our main goal is to show why a
   decelerating universe in gravity time can be treated as an
   accelerating universe in electromagnetic time. In a spatially flat FRW universe filled with dust, the scale factor is

   \be
 a(u)=(u/u_0)^{2/3},
 \label{scale}
   \en

where $a$ is scale factor, which is  shared by electromagnetic time
  and gravity time, $u$ is gravity time, $u_0$ is the
gravity age of the universe and the scale factor has been scaled to
$1$ at the present epoch. Clearly, in the view of gravity time $u$,
the
 universe decelerates all the time. If the gravity time $u$ is a
linear function of the electromagnetic time $t$, we can always
rescale $u$ to make $u$ identify with $t$. It is only a trivial case
 in which the universe decelerates in view of electromagnetic time all the same.
    Giving a most simple nonlinear relation of gravity time $u$ and
electromagnetic time $t$ as follows,
  \be
 u=t^2,
 \label{nonla}
 \en
we obtain deceleration parameter according to electromagnetic time
$t$,
\be
 \frac{1}{a}\frac{d^2a}{dt^2}=\frac{4}{9}t^{-2}.
 \label{decet}
 \en
 From (\ref{decet}), the universe becomes
  accelerating in view of electromagnetic time in this naive toy model (\ref{nonla}), which may be helpful
  to explain the marvelous acceleration of our universe \cite{acce}.

  Note that all our clocks are electromagnetic systems or other gauge
  interaction systems. A mechanical clock with pendulum may be treated
  as a periodic gravity system. In fact, the pendulum swings for a
 long time mainly because of the unwinding of mainspring, not the
gravity. So it must also be taken for electromagnetic system. In a
word, ``time" we used in all the measurements and observations by a
modern clock is electromagnetic time (we assume other gauge
interactions share the same time with electromagnetic interaction).
If we are aware of gravity has an inherent different time concept,
the accelerating universe is easily understood.

  Now we turn to a more realistic multi-time model to simulate $\Lambda$CDM.
  The spatially flat $\Lambda$CDM model becomes the new standard model of cosmology since it well fits the present observation
  data, especially the luminous distance data of type Ia supernovae.
  However, the amplitude of the cosmological constant $\Lambda$ gravely deviates
  from its theoretical predication, approaching the order of
  $10^{122}$. This deviation is accredited as the most serious
  deviation between theory and observation in the history of
  science. Such a severe problem indicates that cosmological
  constant may be only an illusion.

  We have several strong evidences that dark matter exists in our universe,
  whose partition is much larger than that of ordinary matter (baryon
  matter). The standard cold dark matter model (SCDM) supposes that our universe is spatially flat
  and filled with dust. All the expansion data can be explained in  such a
  dust universe with a specific relation of gravity time and
  electromagnetic time. Especially, all the expansion facts which can be
  described by $\Lambda$CDM exactly re-present in SCDM
  with the following relations,

  \be
  \hat{u}=\frac{1}{\hat{r}}\sinh \hat{r}\hat{t}.
  \label{rel scdm}
  \en
  Here $\hat{r}=\Omega_{\Lambda}^{1/2}(\Omega_{mt}\Omega_{mu})^{-1/2},~ \hat{u}=\bar{u}\sqrt{\Omega_{mu}},
  ~\hat{t}=\bar{t}\sqrt{\Omega_{mt}}$~.

 It is better to make some explanations of the above equations and symbols. First of all, we
  have two scenarios of the universe. The traditional scenario (TS) is that all the laws of nature share the same time, in which $\Lambda$CDM's prediction is well consistent with observations.
  Multi-time scenario (MTS) is that gravity and electromagnetic fields cherish their own times,
  respectively. $\Omega_{mt}$ and $\Omega_{\Lambda}$ are the present
  partitions of dust and cosmological constant in TS, respectively.  $\Omega_{mu}$ and
  $\Omega_{K}$ are the present partitions of dust and curvature in
  MTS, respectively. $\bar{t}$ and $\bar{u}$ are the scaled
  electromagnetic time and gravity time in MTS,
  $\bar{t}=H_{0t}t,~\bar{u}=H_{ou}u$.   The electromagnetic time $t$
  in MTS is identical to the universal time in TS. Hence we use the same mark $t$ to denote both the
  electromagnetic time in MTS and the universal time in TS. $H_{0t},~H_{0u}$
  are the present Hubble constants defined by electromagnetic time
  and gravity time, respectively. In cosmology, the scale factor is a freely adjusted quantity up to a
  constant factor. Here we just set that the gravity time and electromagnetic
  time share the same scale factor, ie, $a_u=a_t=a$.

 From (\ref{rel scdm}), one sees that $\hat{u}$ is identical to $\hat{t}$ in the very early universe while $\hat{u}\sim e^{\hat{t}}$ in the late time universe.
 This is to say that gravity time flows at the same rate (up to a constant factor) as electromagnetic time in the early universe but
 flows much  faster than electromagnetic time in the late time universe. So in late time universe, when a supernova recesses
 a distance the gravity time it experienced is more longer than the electromagnetic time it experienced, which yields the fact that the supernova
 looks much  swifter in view of electromagnetic time. This is the essence of the present apparent acceleration in MTS.

   Now we would like to concern more technic details for our real universe. More precisely, some evidences imply that the dark matter together with the luminous matter still cannot flat the universe.
   So in MTS there will be no enough matter to flat the
  universe, whose spatial sector becomes hyperbolic (KCDM). In this case, all the expansion facts which can be
  described by $\Lambda$CDM also exactly re-present in a hyperbolic
  universe with the following relations,

  \be
  \bar{t}=\frac{2}{3\sqrt{\Omega_K}}
  \ln\left[\sqrt{1+r^2\left(\frac{\cosh(\sqrt{\Omega_K}~
  \phi)-1}{2}\right)^3}+r\left(\frac{\cosh(\sqrt{\Omega_K}~
  \phi)-1}{2}\right)^{3/2}\right],
  \label{tbar}
   \en

    \be
  \bar{u}=\frac{\Omega_{mu}}{2{\Omega_K}}
   \left[\frac{1}{\sqrt{\Omega_K}}\sinh(\sqrt{\Omega_K}~\phi)-\phi\right],
   \label{ubar}
   \en

   \be
  r=\frac{\Omega_\Lambda\Omega_{mu}^3}{\Omega_{mt}\Omega_{K}^3}.
   \en

   Here $\Omega_{K}$ is the present partition of curvature in
  MTS. $\bar{t}$ and $\bar{u}$ are the scaled
 electromagnetic time and gravity time, which are implicitly related by a medium parameter $\phi$ in (\ref{tbar}) and (\ref{ubar}).

  As we have mentioned, all of our tools to measure time in
  scientific experiments (observations) is electromagnetic time. We
  have no idea to measure the gravity time in cosmology directly up to now, which yields a great difficulty to
  distinguish MTS from TS by expansion facts. If we cannot tell a
  prediction which is different in MTS and TS, the MTS is only a
  superfluous try.  Fortunately, we find that MTS really tells some
  interesting predictions on the velocity of gravity (wave) and
  equivalence principle.

\begin{figure}
 \centering
 \includegraphics[totalheight=2.5in]{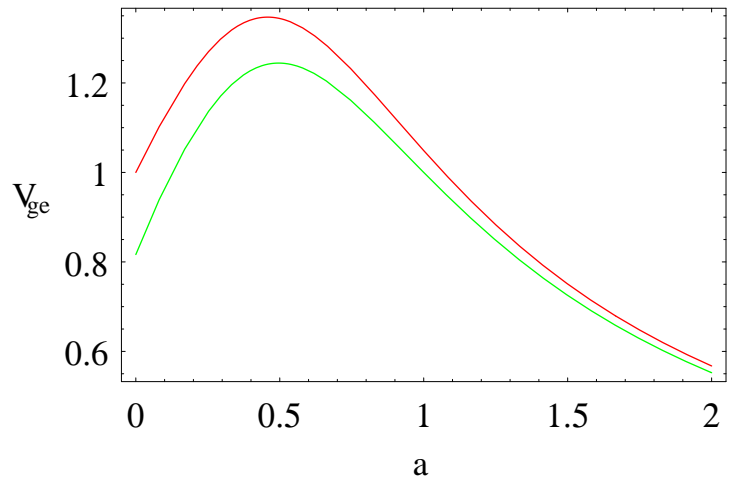}
 \caption{The gravity velocities measured by electromagnetic time as functions of the scale factor.  The two parameter sets, I. $\Omega_{mu} = 0.3, \Omega_{mt} = 0.3$, and II.
 $\Omega_{mu}=0.2, \Omega_{mt}=0.3$ reside on the red and green curves, respectively.}
 \label{velo}
 \end{figure}

  The gravity (wave) propagates at a different velocity measured by electromagnetic time from that of
  light in MTS, if we suppose that  the gravity (wave) propagates exactly
  at the light velocity measured by gravity time. First we point a gauge freedom of $u$. For a constant $b$, if we set $u\to bu$ and replace
  $H_{0u}$ by $b^{-1}H_{0u}$, nothing changes in the above discussions. Thus, without losing any universality, we set
  $H_{0u}=H_{0t}$, that is, the gravity time and electromagnetic time share the same $present$ Hubble parameter. In a small space interval
  $ds$, the gravity (wave) velocity in gravity time $V_g$ relative to what in electromagnetic time $V_e$ reads,
  \be
  V_{ge}\equiv\frac{V_g}{V_e}=\frac{{ds}/{du}}{{ds}/{dt}}=\frac{d\bar{t}}{d\bar{u}},
    \en
  where we have used $H_{0u}=H_{0t}$.  In the case of perfect simulation of $\Lambda$CDM by KCDM in frame of MTS, we obtain
  \be
   V_{ge}=\left(\frac{\Omega_{mu}a^{-3}+\Omega_Ka^{-2}}{\Omega_{mt}a^{-3}+\Omega_{\Lambda}}\right)^{1/2}.
  \en
   So measured by electromagnetic time, the gravity (wave) velocity is a variable in the history of the universe. Clearly, $V_{ge}$ goes to
     $(\Omega_{mu}/\Omega_{mt})^{1/2}$ in the very early universe while it becomes zero in the far future. Fig \ref{velo} illuminates the evolution
     of $V_{ge}$ in the history of the universe.

 Therefore, a certain prediction of MTS is that the gravity velocity in the history of the universe is variable. Some evidences imply that the present velocity of gravity is approximately equal to the velocity of light \cite{gravelo}.  However, its uncertainty is still too big to constrain $\Omega_{mu}$ and $\Omega_{mt}$. The gravity imprints its track of propagation in the early universe. Also the structure formation process will be different
  if gravity and light travel at different velocities. We expect that we can find some clues of the velocity of gravity in the history of the universe in cosmic microwave background and the large scale structure.

 At last, we say some words on the framework of the theory. First, we have two time coordinates in MTS in an FRW universe. A natural proposition
 is that the gravity time together with the three space coordinates obeys the general covariance principle if we
 want to inherit as much achievements of general relativity as possible. The electromagnetic time is a parameter which does not directly construct the 4-geometry of the space-time.
   In a chart other than FRW, the relation between the gravity time and electromagnetic time will formally holds, that is, it is still (\ref{rel scdm}) or (\ref{tbar}) and (\ref{ubar}). Under this assumption,  electromagnetic time is a gravity-induced quantity.     Second, we assume that  the electromagnetic time and gravity time are identified up to a constant factor in a stationary time (with a time-like Killing vector). So we can still enjoy the splendid successes of special relativity and general relativity in the Solar system.

 Now we conclude this letter. We analyzed in detail that the different interactions share the same time concept is not an axiomatic conclusion. We put forward that the present
 cosmic acceleration is natural if gravity and electromagnetic fields do not share a unique time (MTS), without introducing any additional exotic matters or any additional structures of the 4-spacetime. Furthermore, we present a   certain testable predication of MTS, that is, the velocity of gravity is a variable in the history of the universe. Our future works include: To verify MTS or defy it by observation, to explore it by fundamental theory, ie, to present a fundamental demonstration of the relation of electromagnetic time and gravity time, not only put it phenomenologically etc.


    \vskip 0.5cm

 {\bf Acknowledgments.} The author thanks Y  Xu for preparing fig 1 and Y Du for carefully editing the early version of this letter.

\end{document}